\documentclass[fleqn,twoside]{article}
\usepackage{espcrc2}

\usepackage{graphicx}

\newcommand{\AmS}{{\protect\the\textfont2
  A\kern-.1667em\lower.5ex\hbox{M}\kern-.125emS}}

\title{Localization and lattice fermions\thanks{Presented at the Workshop on Computational Hadron Physics, Nicosia, Cyprus, September 14--17, 2005.}}

\author{Benjamin Svetitsky\address{School of Physics and Astronomy, Raymond and Beverly Sackler Faculty of Exact Sciences,\\ Tel Aviv University, 69978 Tel Aviv, Israel}%
        }
       
\begin{document}

\begin{abstract}
I review how the phenomenology of localization applies to fermions in lattice gauge theory, present measurements of the localization length and other quantities, and discuss the consequences for things like the overlap kernel.
\vspace{1pc}
\end{abstract}

\maketitle

Localization is a phenomenon long studied in condensed-matter physics
\cite{Thouless}.  Concerned with conduction in disordered media, it is based on the study of eigenstates of the Schr\"odinger equation in a random potential.  Its counterpart in lattice gauge theory is the study of the spectrum and eigenstates of a Hermitian fermion kernel in the fluctuating gauge field.  Here we deal with Wilson fermions, with kernel $H_W=\gamma_5D_W$, in an assortment of gauge ensembles \cite{GS,GSS}.  We find low-lying {\em localized\/} states, extending up to an energy called the {\em mobility edge\/}; and {\em extended\/} states at higher energy.  We characterize the localized states by their {\em localization length\/} and their {\em support length} (related to their {\em inverse participation ratio}, or IPR).
In this talk I will expand a bit on our methods for calculating these quantities, which are based on calculations of fermionic Green functions.  

The Wilson fermion kernel is not so popular these days for its own sake but rather for its role in constructing domain-wall fermions~\cite{DW} and the closely related overlap fermions~\cite{NN}.
The simplest overlap kernel is given by
\begin{equation}
D_{\textrm{ov}}=1-\gamma_5\,\textrm{sgn}(H_W).\label{DOV}
\end{equation}
The range of $D_{\textrm{ov}}$ can become long if the spectrum of $H_W$ contains low-lying (or zero) modes~\cite{HJL}.
Whether this happens, however, depends on whether the low-lying modes are localized.
Localization is reflected in the Green functions of the theory.
On the one hand, this enables us to use Green functions to calculate localization quantities; on the other hand, the range of the overlap kernel, which is just a Green function, will be influenced directly by these quantities.
We will argue here that, though the energy threshhold of $H_W$ lie at zero, the range of $D_{\textrm{ov}}$ is determined by the position of the mobility edge---that is, the low-lying modes are generally harmless.
The mobility edge perforce lies at non-zero energy, as long as one avoids the Aoki phase~\cite{Aoki}.  Indeed, the descent of the mobility edge to zero serves as a useful definition of the onset of the Aoki phase.

\section{Localization basics}

\begin{figure*}[htb]
\begin{center}
\includegraphics*[width=1.4\columnwidth]{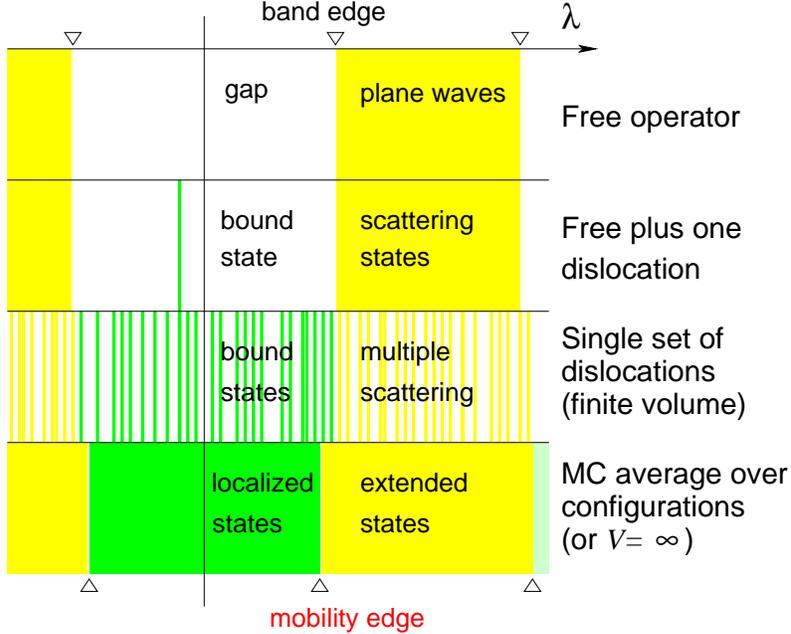}
\caption{How the spectrum of $H_W$ changes as one adds (top to bottom) one dislocation, many dislocations, and a random ensemble of dislocations to the gauge field.}
\label{fig1}
\end{center}\end{figure*}

The Wilson fermion kernel is
\begin{equation}
   D_W(m_0) =  {1\over a}\left(\begin{array}{cc}
      (W + a m_0)   & -C     \\
      C^\dag  & (W + a m_0)
       \end{array}\right).
\end{equation}
Here
\begin{equation}
C_{xy}  =  \frac12 \sum_\mu \left[\delta_{x+\hat\mu,y} U_{x\mu}
    - \delta_{x-\hat\mu,y} U^\dag_{y\mu} \right] \sigma_\mu
\end{equation}
is the naive Dirac term while
\begin{equation}
W_{xy} = 4\delta_{xy} -\frac12 \sum_\mu \left[\delta_{x+\hat\mu,y} U_{x\mu}
               + \delta_{x-\hat\mu,y} U^\dag_{y\mu} \right]
\end{equation}
is the Wilson term.
A Hermitian operator can be obtained by defining
\begin{equation}
H_W=D_W\gamma_5=H_W^\dag,
\end{equation}
and it is clear that $H_W$ is a ``Hamiltonian'' of a fermion moving
in the random gauge field $U_{x\mu}$.  Dislocations in any given configuration $U_{x\mu}$ can create bound states, and even zero modes \cite{BNN}, in the spectrum of $H_W$.  Of course we do not study single configurations of $U_{x\mu}$ but averages over an ensemble. The result is localization. 

For a general overview, let's look at Fig.~\ref{fig1}, which shows how localization can appear as we go from free fermions (i.e., fixing
$U_{x\mu}=1$) to the fully fluctuating gauge field.
We assume that the free operator $H_W$ has a gap for the parameters we choose; above that gap lie plane-wave states.  Condensed-matter physicists call the top of the gap the band edge. If we add a single dislocation to the otherwise constant gauge field, the plane waves will be replaced by scattering states and a bound state might appear in the gap.  A large number of dislocations will create a large number of bound states, and the scattering states will show effects of multiple scattering.  In finite volume, the scattering states as well as the bound states form a discrete spectrum.

In the infinite volume limit, with a fixed density of dislocations, both the bound states and the scattering states form a continuum.  The energy that separates the two is called the mobility edge.  The infinite volume limit automatically gives us an average over the shapes and positions of the dislocations in any finite subvolume.  Alternatively, we can keep the volume finite and average over gauge field configurations ourselves, as is done in lattice gauge theory.

$H_W$ can only have zero eigenvalues in the supercritical region, $-8<m_0<0$.  As it happens, this includes the region of interest for the construction of domain-wall and overlap fermions.  The Aoki phase lies in this supercritical region; we stay outside the Aoki phase so that the mobility edge is above zero.  We will present numbers \cite{GSS} below for $m_0=-1.5$.  For the gauge couplings we choose, the theory lies between the Aoki ``fingers.'' These couplings are in fact popular among users of domain-wall and overlap fermions.

\section{The spectral density}

Since averaging over the gauge field smears the energy eigenvalues, there is never a discrete spectrum of bound states accompanied by a continuum of extended states.  The spectrum in the disordered system is described by a {\em continuous} density of states,
\begin{equation}
\rho(\lambda)= {1\over V} \Big\langle\sum_n \delta(\lambda-\lambda_n)\Big\rangle,
\end{equation}
where $\lambda$ is the energy eigenvalue. 
\begin{figure}[htb]
\includegraphics*[width=.9\columnwidth]{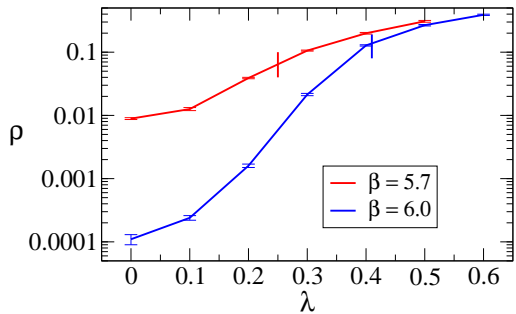}
\caption{Eigenvalue density $\rho(\lambda)$, in lattice units, for $\beta=5.7$ ($a^{-1}=1$~GeV, upper curve) and $\beta=6.0$ ($a^{-1}=2$~GeV, lower curve).  Plaquette action.}
\label{fig2}
\end{figure}
We show in Fig.~\ref{fig2} the density of states for the gauge ensemble generated with the Wilson plaquette action.
The vertical bars indicate the mobility edge for the two couplings.
Nothing special happens at the mobility edge.

[We find it convenient to calculate $\rho$ via the identity
\begin{equation}
\pi\rho(\lambda) 
  = {1\over V}\lim_{m_1\to 0} \langle{{\rm Im}\, {\rm Tr }\,G(\lambda+im_1)}\rangle, 
  \end{equation}
where $G(z)=(H_W-z)^{-1}$ is the resolvent of $H_W$.  The mobility edge is where the averaged localization length $l_\ell(\lambda)$ (see below) diverges.]

The main point of Fig.~\ref{fig2} is to show that there is indeed a nonzero density of states at zero energy.
Table~\ref{table1} shows how this density changes with cutoff and with the choice of gauge action.  Improved actions dramatically lower $\rho(0)$, but in no case is it actually zero.
\begin{table}
\caption{Eigenvalue density $\rho(0)$ at $\lambda=0$ for three gauge
actions, each at two values of the gauge coupling.\label{table1}}
\begin{tabular}{cccc}
\hline
Cutoff $a^{-1}$&Wilson&Iwasaki&DBW2\\\hline
1~GeV&$10^{-2}$&$6\times10^{-3}$&$10^{-3}$\\
2~GeV&$10^{-4}$&$7\times10^{-7}$&$<10^{-7}$\\
\hline
\end{tabular}
\end{table}

In a theory with dynamical Wilson fermions, these zero modes will be of no consequence because they will cause the fermion determinant to vanish.  
In any other theory, however, the spectrum of $H_W$ does not directly determine the fermion determinant.  Whether the ensemble is quenched, or whether it contains the domain-wall or overlap fermion determinant,
the ensemble average will furnish $H_W$ with modes near (and at) zero energy.
Will this destroy the locality of $D_{\textrm{ov}}$?
No---because (outside the Aoki phase) these modes are localized.%
\footnote{This is not to say that these modes won't cause any trouble at all.  Low-lying modes give $H_W$ a bad condition number, and make it difficult to calculate the overlap kernel via various approximations to
Eq.~(\ref{DOV}).  See for instance Refs.~\cite{Degrand,Duerr}.} 

\section{Localized modes}

\begin{figure*}[htb]
\begin{center}
\includegraphics*[width=.6\textwidth]{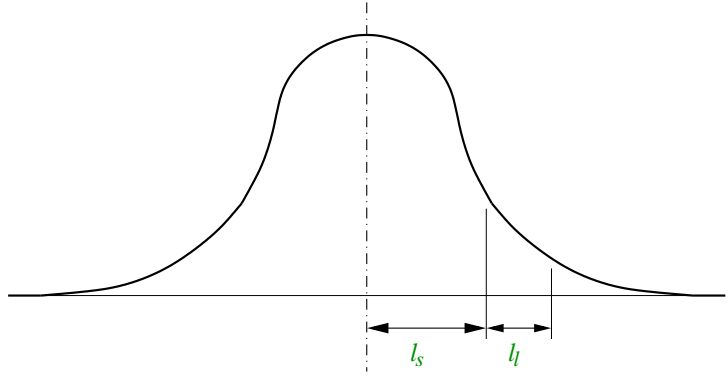}
\caption{Characteristic lengths of a localized eigenfunction: the support length $l_s$, which is the size of the region containing most of the amplitude; and the localization length $l_\ell$, which is the decay length of the tail.}
\end{center}
\label{fig3}
\end{figure*}

A localized mode can be described by its support length $l_s$, which is the size of the region that contains most of the mode's density; and by its localization length $l_\ell$, which is the decay length of any long-range tail (see Fig.~\ref{fig3}).  Each can be averaged over all eigenmodes with eigenvalue $\lambda$, giving $l_\ell(\lambda)$ and $l_s(\lambda)$.

The average support length can be used to determine in what range of $\lambda$ the localized modes are dilute.  In view of the mode density shown in Fig.~\ref{fig2}, it is clear that low-lying modes will be dilute while the higher modes will be more crowded.
The average localization length, on the other hand, will directly affect the decay rate of Green functions of $H_W$.
We turn \cite{GSS} to these Green functions to calculate both $l_\ell(\lambda)$ and $l_s(\lambda)$.

We define a ``meson'' Green function as a pair propagator,
\begin{equation}
\Gamma(x,y)=\left(\frac1{H-z}\right)_{xy}\left(\frac1{H-z^*}\right)_{yx},
\end{equation}
where the complex variable $z=\lambda+im_1$.  Introduction of $\lambda$ will enable us to focus on the contributions of eigenfunctions with eigenvalue near $\lambda$; the imaginary part $m_1$ is a regulator that is put in to avoid the singularities on the real axis due to these eigenfunctions.

The spectral representation of $\Gamma$ is a double sum,
\begin{eqnarray}
 \Gamma(x,y)&=&
  \Bigl\langle
    \sum_{n^{\pm}}
    \Psi_{n^+}^\dagger(x) \Psi_{n^-}(x) {1\over \lambda_{n^-} - \lambda + im_1}\nonumber\\
  &&\times  \Psi_{n^-}^\dagger(y) \Psi_{n^+}(y) {1\over \lambda_{n^+} - \lambda - im_1}
  \Bigr\rangle.\nonumber\\
\end{eqnarray}
In the limit $m_1\to0$ it diverges as $1/m_1$; the divergence comes from the terms where $n^+=n^-$, and thus we are left with a single sum,
\begin{eqnarray}
  \Gamma(x,y)& =& {1\over m_1}
  \biggl\langle
    \sum_n |\Psi_n(x)|^2 |\Psi_n(y)|^2 \nonumber\\
    &&\times{m_1 \over (\lambda_n - \lambda)^2 + m_1^2}
  \biggr\rangle + O(1).
\end{eqnarray}
The $m_1\to0$ limit has thus brought out the eigenmode densities
$|\Psi_n(x)|^2$.
The limit also focuses on modes with $\lambda_n\approx\lambda$, since the fraction within the bracket approaches a delta function as $m_1\to0$.

Now we set $x=0$, $y=({\bf y},t)$, and sum over the spatial coordinate {\bf y}.  This gives the correlation function $\Gamma(t)$ between spatial slices.  Taking finally the $m_1\to0$ limit, we find
\begin{equation}
  \lim_{m_1 \to 0} m_1 \Gamma(t)
  \sim \pi \rho(\lambda) \int dl\, {\cal P}_\lambda(l)\,
e^{-t/l},
\end{equation}
where ${\cal P}_\lambda(l)$ is the distribution of localization lengths among the modes with eigenvalue $\lambda$.
The tail of $\Gamma(t)$ thus gives an average of the tails of the mode densities $|\Psi_n(x)|^2$, and the average 
$\langle l\rangle_\lambda$ can serve as an estimate of the localization length $l_\ell(\lambda)$.
By definition, $l_\ell(\lambda)$ diverges at $\lambda=\lambda_c$,
the mobility edge.
Clearly this happens when the modes at $\lambda$ become extended, meaning they have spread across the lattice.

The same tools give us an estimate of the average support length
$l_s(\lambda)$.
If a given (normalized) mode density $|\Psi_n(x)|^2$ takes a value $\approx1/l_s^d$ over a region
of linear size $l_s$, then
\begin{eqnarray}
  m_1 \Gamma(t=0) &=&
  {1\over V} \biggl\langle{\sum_n
    {1\over l_{s}(n)}\,
    {m_1 \over (\lambda_n - \lambda)^2 + m_1^2}}\biggr\rangle\nonumber\\[3pt]&&+O(m_1),
\end{eqnarray}
which approaches
$$    \pi\rho(\lambda)\frac1{l_s(\lambda)} $$
as $m_1\to0$.
Thus we obtain an estimate of the average support length.
Comparing $l_s(\lambda)$ to $\rho(\lambda)^{-1/3}$, which is the mean distance between localized modes, tells us whether the modes at $\lambda$ overlap with each other or whether they are dilute.

\section{Back to the overlap kernel}

The dilute, the less dilute, and the extended modes all contribute to the overlap kernel $D_{\textrm{ov}}$.  The contribution of the dilute localized modes, those below some energy $\bar\lambda$, is of the form
\begin{eqnarray}  \langle{|D_{\textrm{ov}}(x,y)|}\rangle_{\textrm{loc}}
  &\approx& \int_{-\bar\lambda}^{\bar\lambda} d\lambda\, \rho(\lambda)
  \nonumber\\&&\qquad\times\exp\left(-{|x-y|\over 2l_\ell(\lambda)}\right)\nonumber\\[3pt]
  &\approx& \bar\lambda \rho(\bar\lambda)
  \exp\left(-{|x-y|\over 2l_\ell(\bar\lambda)}\right),\nonumber\\
\end{eqnarray}
where the second line follows from the steep rise in $\rho(\lambda)$ seen in Fig.~\ref{fig2}.  Note that these isolated modes contribute an exponential tail characterized by their localization length; their low energy is harmless.
The dense modes are up near the mobility edge $\lambda_c$; if we lump them with the extended modes then we can estimate that the fall-off of their contribution will be governed by $\lambda_c$, viz.
\begin{equation}
  \langle{|D_{\textrm{ov}}(x,y)|}\rangle_{\textrm{ext}}
  \approx {\cal C} \exp\left(-\lambda_c |x-y|\right),
\end{equation}
where $\cal C$ is $O(1)$.

Let's put in numbers for the plaquette action at $\beta=6.0$.  We measure the mobility edge to be at $\lambda_c\simeq0.41$ and we fix the demarcation point to be $\bar\lambda\approx0.2$.  The result is
\begin{eqnarray}
  \langle{|D_{\textrm{ov}}(x,y)|}\rangle
  &\approx& {10^{-4}} \exp\left(-{|x-y|\over {1.4}}\right)
  \nonumber\\[3pt]
  &&+{{\cal O}(1)}\cdot \exp\left(-{|x-y|\over {2.4}}\right).
  \label{range}
\end{eqnarray}
The second term, governed by the mobility edge, wins---both in prefactor and in exponent.  We find this to be true in all the cases (actions and couplings) we have studied, including those in Table 1.
The reverse could be true in other cases; then the range of $D_{\textrm{ov}}$ would be determined by the representative localization length, $l_\ell(\bar\lambda)$.  The fact that the gap is zero would still be harmless.

We can interpret $\lambda_c$ as a mass scale, the mass of effective excitations that influence the overlap kernel.  When the cutoff is $a^{-1}=2$~GeV, we find that $\lambda_ca^{-1}\simeq800$~MeV for all three actions.  It may be a disappointment to find that the cutoff is felt at an energy that is so low, but this is still well above $\Lambda_{\rm QCD}$ which is the true dynamical scale. One ought to worry, however, at stronger couplings.  When $a^{-1}=1$~GeV, the value of $\lambda_ca^{-1}$ comes out to be only 250--320~MeV, not a place we would like to see unphysical particles in the spectrum.

\section{Conclusions}

The three main lessons of our work:
\begin{enumerate}
\item
The mobility edge at $\lambda_c>0$ [or $l_\ell(\bar\lambda)$, if it is greater than $\lambda_c^{-1}$] assures a finite range for $D_{\textrm{ov}}$, even though $H_W$ has no gap in the disordered
gauge field.
\item
One should demand that $\lambda_ca^{-1}\gg\Lambda_{\rm{QCD}}$.
\item
$\lambda_ca^{-1}$ is fairly insensitive to the gauge action (at fixed cutoff), even as $\rho(0)$ varies widely.
\end{enumerate}

The third point emerges from numerical results \cite{GSS} that are not presented here; it implies, according to our arguments, that the range of the overlap kernel will be equally insensitive to the gauge action.
This has already been reported in Ref.~\cite{Degrand}; moreover, the range of $D_{\textrm{ov}}$ given there for $a^{-1}\approx 2$~GeV agrees well with our estimate, shown in Eq.~(\ref{range}) above.
It is curious that the range given for the overlap kernel when $a^{-1}\approx 1$~GeV in Ref.~\cite{Draper} is much shorter (by a factor of 4) than our estimate would indicate.  The reason for this is an interesting open question.

I have not discussed the implications of localization for domain-wall fermions, where it would govern both the range of the equivalent overlap-like kernel and the residual mass due to finiteness of the fifth dimension in practical calculations \cite{GS,GSS}.  Numerical results from QCDOC were reported by Peter Boyle at {\em Lattice 2005} \cite{Boyle}.

\section*{Acknowledgments}
This work was supported by the Israel Science Foundation under
grant no.~222/02, the Basic Research Fund of Tel Aviv
University, and the US Department of Energy.
Our computer code is based on the public lattice gauge theory code
of the MILC collaboration.
The code was run on supercomputers of the Israel Inter-University Computation Center and on  a Beowulf cluster at San Francisco State University.

\end{document}